\documentclass [12pt,titlepage]{article}
\usepackage {graphicx}
\title{ANOMALOUS MAGNETIC MOMENT AND LAMB SHIFT IN THE SECOND ORDER OF THE
PERTURBATION THEORY\\ WITH FINITE ELECTRON MASS RENORMALIZATION IN
QUANTUM ELECTRODYNAMICS}
\author{V.P. Neznamov}
\begin{document}
\maketitle
\begin{abstract}
The paper determines the anomalous magnetic moment and Lamb energy
level shift in the second order of the perturbation theory using
the algorithm of self-energy expression regularization in quantum
electrodynamics (QED) that meets the relativistic and gauge
invariance requirements /1/; the comparison is made to the
associated conventional quantum electrodynamics results.
 A limiting 4-impulse, $L^2 = L_0^2 - \vec {L}^2$,
with an infinitely large temporal component ($\frac{L_0 }{m} > >
1)$ and limited value of spatial components $L_{i}$ that depend
quite weakly on the Dirac particle impulse variation is introduced
within the algorithm.

The computations in the second order of the perturbation theory
result in the anomalous magnetic moment of electron, which is
larger by $\sim 2.4\,$\% than that in the conventional QED
calculations, and the Lamb hydrogen level shift $\Delta
v_{2S_{1/2}}  - \Delta v_{2P_{1/2}}  $, which is smaller by $\sim
6\,$ \% than that in the conventional QED calculations.

The answer in regard to the agreement between the experimental
data and results of the calculations by the algorithm of this
paper will be given by the calculations of the next order of the
perturbation theory which are being planned for the nearest
future.

Merits of the algorithm suggested are:
\begin{itemize}
  \item  possibility of finite renormalization of electron and positron
mass
\[
\left( {m^{\left( {2} \right)} = m_{0} \left( {1 +
1.115\frac{{\alpha }}{{\pi} }} \right)} \right);
\]
  \item  agreement between the final results for the self-energy in the
second order of the "old" and relativistically covariant
perturbation theory /1/;
  \item provision of the symmetric interaction of the electron intrinsic
magnetic moment with the external magnetic and electric fields
(the electron interacts both with the external magnetic field and
external electric field through its intrinsic magnetic moment in
the same manner in the first and second order of the perturbation
theory).
\end{itemize}
\end{abstract}
\tableofcontents %\contentsname
\newpage
%\subsubsection{TABLE OF CONTENTS}

% page
%1. Introduction 4
%2. Anomalous magnetic moment 7
%Vertex operator 7
%Mass operator. Mass renormalization 9
%3. Lamb shift 10
%3.1. Vertex operator 10
%3.2. Mass operator. Mass renormalization 13
%3.3. Polarization operator. Charge renormalization 14
%4. Discussion of results 16
%4.1. Anomalous magnetic moment 17
%4.2. Lamb atomic energy level shift 17
%References21

\section{Introduction}

Ref. /1/ suggests an algorithm for regularization of the
self-energy expressions for Dirac particle that meets the
relativistic and gauge invariance requirements. A limiting
4-impulse, $L^2 = L_0^2 - \vec {L}^2$, with an infinitely large
temporal component ($\frac{L_0 }{m} > > 1)$ and limited value of
spatial components $L_{i}$ depending quite weakly on the particle
impulse variation is introduced within the algorithm. For the
particle at rest, $ | \vec {L} | = m$; for the particle impulses
$\frac{p}{m}$ ranging from 0 to 1, $ | \vec {L} | \simeq m\left(
{1 + 0.202\frac{p^{\;4}}{m^{\;4}} - 0.135\frac{p^6}{m^6}}
\right)$; for the ultrarelativistic case, $ | \vec {p} | > > m: |
\vec {L} | \approx 2m$.

The dependence of the components $L_{i}$ on the particle impulse
implies the relevant dependence on the impulse of the temporal
component $L_{0}$ ensuring invariant $L^2 = C m^2$, where $C$ is a
numerical coefficient much larger than one.

In the presence of external fields $A^\mu \left( \vec {x} \right)$ the
components $L_{0}$, $L_{i}$ depend on their magnitude with the invariant
$L^{2}$ remaining unchanged.

Within the algorithm suggested, in the second order of the perturbation
theory the renormalized mass of Dirac particle is

\[m^{(2)} = m_0 + \Delta m^{(2)} = m_0 \left( {1 +
1.115\frac{\alpha }{\pi }} \right),\] where $m_{0}$ is the
``bare'' mass of the particle, $\alpha $ is the fine-structure
constant.

This paper uses the algorithm /1/ to determine the anomalous magnetic moment
and Lamb energy level shift in the second order of the perturbation theory.
The contribution of Feynman diagrams $1a\div 1f$ presented in Fig.1 to the
physical processes is calculated specifically.

\begin{figure}[htbp]
  \begin{center}
    \includegraphics[width=11.8cm,keepaspectratio]{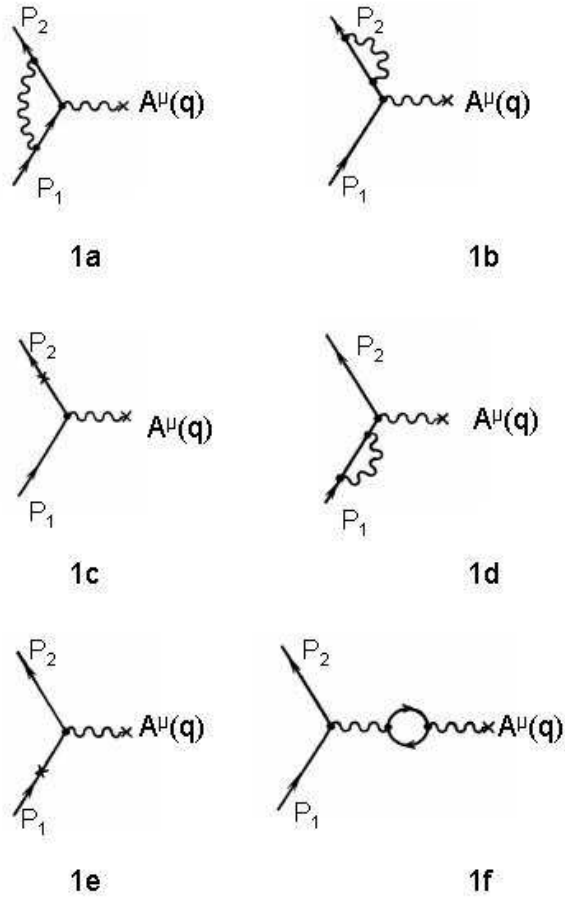}
    \caption{Electron scattering in external field $A^{\mu }(q)$.
 Diagrams of the second order of the perturbation theory.}
    \label{fig_1}
  \end{center}
\end{figure}
%\begin{figure*}[p]
%\centering
%\includegraphics[width=118mm,height=175mm]{Pic.jpg}
%\caption{Electron scattering in external field $A^{\mu }(q)$.
% Diagrams of the second order of the perturbation theory}
%\end{figure*}

 In so doing the case of non-relativistic motion of charged
particle $\left( {|\vec {p}|< < m} \right)$ in weak external
static magnetic or electric field is considered.

Like in /1/, the following computational steps are performed:
\begin{enumerate}
  \item The integration with respect to variable $k^{0}$ (for the vertex and
mass operators) or variable $p^{0}$ (for the polarization
operator) is performed using the residue theorem and Feynman rule
of pole bypassing.
  \item The integration of impulse $k$ (for the vertex and mass
operators) with respect to spatial variables is performed with
introduction of the finite limit of integration $|\vec {L}|= m$.
\end{enumerate}

Section 2 calculates the radiation corrections for the electron
motion in a weak static homogeneous magnetic field.

Section 3 calculates the radiation corrections for the electron motion in a
weak static electric field.

Section 4 summarizes the calculations and compares them to the associated
results of the conventional quantum electrodynamics (QED).

\section{Anomalous magnetic moment}

Consider the electron motion in a weak external homogeneous magnetic field
$\vec {H}$. In the final expressions we restrict our consideration only to
the terms proportional to electron impulse $p_{i}$ and magnetic field
$H_{i}$. Expressions proportional to $p_{i}H_{k}$ will be neglected. A
contribution to the anomalous magnetic moment is made only by diagrams
$1a\div 1e$.

\subsection{Vertex operator}

Initially consider the contribution to the anomalous magnetic
moment made by the diagram $1,a$ with vertex operator $\Lambda
_\mu ^{(3)} \,(p_1 ,p_2 )$. Write the process amplitude as $e
\cdot \bar {u}(p_2 ) \cdot \Lambda _\mu ^{(3)} (p_1 ,p_2 ) \cdot
A^\mu \cdot u(p_1 )$, where $u(p_{1})$, $\bar {u}(ð_2 )$ are the
relevant bispinors for the initial and final state and $A^\mu $ is
the 4-vector of the external electromagnetic field.

The vertex operator $\Lambda _\mu ^{(3)} (p_1 ,p_2 )$ can be
represented as
\begin{equation}
\label{eq1} \Lambda _\mu ^{(3)} (p_1 ,p_2 ) = \frac{ - i
e^2}{(2\pi )^{\;4}}\int {d^{\;4}k\frac{\left( {4m^2 - 2q^2}
\right) \gamma _\mu - 4m  k_\mu - 2\left( {\hat {q}\gamma _\mu
\hat {k} - \hat {k}\gamma _\mu \hat {q}} \right) - 2\hat {k}\gamma
_\mu \hat {k}}{k^2\left( {\left( {k - p_1 } \right)^2 - m^2}
\right)\left( {\left( {k - p_2 } \right)^2 - m^2} \right)}}.
\end{equation}

In (\ref{eq1}), $q = p_{2} -- p_{1}$.

When the electron moves in weak magnetic field and only those
terms that are proportional to $p_{i}$ and $H_{i}$ are taken into
consideration, the product $e \cdot A^\mu \cdot \Lambda _\mu
^{(3)} (p_1 ,p_2 )$ can be written as
\begin{equation}
\label{eq2}
\begin{array}{l}
 \left( { - e} \right) \cdot A^i \cdot \Lambda ^{(3)\,i}(p_1 ,p_2 ) = \\
 { }^ = \left( { - e} \right)\frac{ - ie^2}{(2\pi )^{\;4}}\int {d^{\;4}k\frac{B + C
\cdot k_0 + D \cdot k_0^2 }{\left( {k_0 - k} \right){\kern 1pt}
\left( {k_0 + k} \right){\kern 1pt} \left( {k_0 - p_1^0 - \tilde
{p}_1^0 } \right){\kern 1pt} \left( {k_0 - p_1^0 + \tilde {p}_1^0
} \right){\kern 1pt} \left( {k_0 - p_2^0 - \tilde {p}_2^0 }
\right){\kern 1pt} \left( {k_0 - p_2^0 + \tilde {p}_2^0 }
\right)}}.
 \end{array}
\end{equation}
In (\ref{eq2}), $B = \left( {4m^2 - 2k^2} \right)\,\vec {\gamma
}{\kern 1pt} \vec {A} + 4\left( {\vec {\gamma }{\kern 1pt} \vec
{k}} \right)\,\left( {\vec {A} \cdot \vec {k}} \right)$; $C = -
4\gamma ^0 \cdot \vec {\sigma }{\kern 1pt} \vec {H}$; $D = 2\vec
{\gamma }{\kern 1pt} \vec {A};$ $\tilde {p}_1^0 = \left( {m^2 +
\left( {\vec {p}_1 - \vec {k}} \right)^2} \right)^{1 / 2};\;\tilde
{p}_2^0 = \left( {m^2 + \left( {\vec {p}_2 - \vec {k}} \right)^2}
\right)^{1 / 2}.
$

Next, following the action program discussed in Introduction,
integrate with respect to variable $k_{\mbox{î}}$ with using the
residue theorem, for example, for the poles in the right
half-plane of the complex variable $k_0$. Then expression
(\ref{eq2}) becomes
\begin{equation}
\label{eq3}
\begin{array}{l}
 \left( { - e} \right) \cdot A^i \cdot \Lambda ^{(3){\kern 1pt} {\kern 1pt}
i}(p_1 ,p_2 ) = \\\quad  = (e)\frac{ie^2}{(2\pi )^{\;4}} \cdot
\left( { - 2\pi i} \right)\int {d\vec {k}\left[ {\frac{B + C \cdot
k + D \cdot k^2}{2k \cdot \left( {k - p_1^0 - \tilde {p}_1^0 }
\right){\kern 1pt} \left( {k - p_1^0 + \tilde {p}_1^0 }
\right){\kern 1pt} \left( {k - p_2^0 - \tilde {p}_2^0 }
\right){\kern 1pt} \left( {k - p_2^0 + \tilde {p}_2^0 } \right)} +
} \right.}\\[7pt]%
\qquad \qquad\quad\qquad+ \frac{B + C\left( {p_1^0 + \tilde
{p}_1^0 } \right) + D\left( {p_1^0 + \tilde {p}_1^0 }
\right)^2}{\left( {\tilde {p}_1^0 + p_1^0 - k} \right)\;\left(
{\tilde {p}_1^0 + p_1^0 + k} \right) \cdot 2\tilde {p}_1^0 \left(
{\tilde {p}_1^0 + p_1^0 - p_2^0 - \tilde {p}_2^0 } \right)\;\left(
{\tilde {p}_1^0 + p_1^0 - p_2^0 + \tilde {p}_2^0 } \right)} +
\\[7pt]
 \qquad \qquad\quad\qquad+ \left. {\frac{B + C\left( {p_2^0 + \tilde {p}_2^0 } \right) + D\left(
{p_2^0 + \tilde {p}_2^0 } \right)^2}{\left( {\tilde {p}_2^0 +
p_2^0 - k} \right)\;\left( {\tilde {p}_2^0 + p_2^0 + k} \right)
\cdot 2\tilde {p}_2^0 \left( {\tilde {p}_2^0 + p_2^0 - p_1^0 -
\tilde {p}_1^0 } \right)\;\left( {\tilde {p}_2^0 + p_2^0 - p_1^0 +
\tilde {p}_1^0 } \right)}} \right].
 \end{array}
\end{equation}

Expression $\left( {\tilde {p}_1^0 + p_1^0 - p_2^0 - \tilde {p}_2^0 }
\right)$ appears in the denominators of the second and third addends in (\ref{eq3}),
and it vanishes with $\vec {p}_1 ,\vec {p}_2 \to 0$. In fact, this feature
is seeming, and it is removed on appropriate algebraic transformations.
After that (\ref{eq3}) can be written as
\begin{equation}
\begin{array}{l}
 \left( { - e} \right) \cdot A^i \cdot \Lambda ^{(3){\kern 1pt}
{\kern 1pt} i}(p_1 ,p_2 ) = \\[7pt]%
 = (e)\frac{e^2}{2 \cdot (2\pi )^3}\,\int {d\vec {k}\left\{
{\frac{\left( {4m^2 - 2k^2} \right)\,\vec {\gamma }{\kern 1pt}
\vec {A} + 4\left( {\vec {\gamma }{\kern 1pt} \vec {k}}
\right)(\vec {A}{\kern 1pt} \vec {k}) - 4\gamma ^0 \cdot \vec
{\sigma }{\kern 1pt} \vec {H} \cdot k + 2 \cdot k^2 \cdot \vec
{\gamma }{\kern 1pt} \vec {A}}{k\left( {\tilde {p}_1^0 - k + p_1^0
} \right)\,\left( {\tilde {p}_1^0 + k - p_1^0 } \right)\,\left(
{\tilde {p}_2^0 - k + p_2^0 } \right)\,\left( {\tilde {p}_2^0 + k
- p_2^0 } \right)} - } \right.}\\[20pt]%
 - \frac{\left[ {\left( {4m^2 - 2k^2} \right)\,\vec {\gamma
}{\kern 1pt} \vec {A} + 4\left( {\vec {\gamma }{\kern 1pt} \vec
{k}} \right)(\vec {A}{\kern 1pt} \vec {k})} \right] \cdot \left[
{\left. {\tilde {p}_2^0 \left( {\left( {\tilde {p}_2^0 + p_2^0 }
\right)^2 - k^2} \right) + \tilde {p}_1^0 \left( {\left( {\tilde
{p}_1^0 + p_1^0 } \right)^2 - k^2} \right) + 2\tilde {p}_1^0 \cdot
\tilde {p}_2^0 \left( {\tilde {p}_1^0 + p_1^0 + \tilde {p}_2^0 +
p_2^0 } \right)} \right]} \right.}{\tilde {p}_1^0 \cdot \tilde
{p}_2^0 \left( {\tilde {p}_1^0 + \tilde {p}_2^0 + p_1^0 - p_2^0 }
\right)\,\left( {\tilde {p}_2^0 + \tilde {p}_1^0 - p_1^0 + p_2^0 }
\right)\,\left( {\left( {\tilde {p}_1^0 + p_1^0 } \right)^2 - k^2}
\right)\,\left( {\left( {\tilde {p}_2^0 + p_2^0 } \right)^2 - k^2}
\right)} + \\[20pt]%
 + \frac{4\gamma ^0 \cdot \vec {\sigma }{\kern 1pt} \vec
{H}\left[ {\tilde {p}_2^0 \left( {\tilde {p}_1^0 + p_1^0 }
\right)\left( {\left( {\tilde {p}_2^0 + p_2^0 } \right)^2 - k^2}
\right) + \tilde {p}_1^0 \left( {\tilde {p}_2^0 + p_2^0 }
\right)\left( {\left( {\tilde {p}_1^0 + p_1^0 } \right)^2 - k^2}
\right) + 2\tilde {p}_1^0 \cdot \tilde {p}_2^0 \left( {\left(
{\tilde {p}_1^0 + p_1^0 } \right){\kern 1pt} \left( {\tilde
{p}_2^0 + p_2^0 } \right) + k^2} \right)} \right]}{\tilde {p}_1^0
\cdot \tilde {p}_2^0 \left( {\tilde {p}_1^0 + \tilde {p}_2^0 +
p_1^0 - p_2^0 } \right){\kern 1pt} \left( {\tilde {p}_2^0 + \tilde
{p}_1^0 - p_1^0 + p_2^0 } \right){\kern 1pt} \left( {\left(
{\tilde {p}_1^0 + p_1^0 } \right)^2 - k^2} \right){\kern 1pt}
\left( {\left( {\tilde
{p}_2^0 + p_2^0 } \right)^2 - k^2} \right)} - \nonumber\\[20pt]%
\left.- \frac{2\vec {\gamma }{\kern 1pt} \vec {A}\left[ {\left(
{\tilde {p}_1^0 + p_1^0 } \right)^2 \cdot \tilde {p}_2^0 \left(
{\left( {\tilde {p}_2^0 + p_2^0 } \right)^2 - k^2} \right) +
\left( {\tilde {p}_2^0 + p_2^0 } \right)^2 \cdot \tilde {p}_1^0
\left( {\left( {\tilde {p}_1^0 + p_1^0 } \right)^2 - k^2} \right)
+ 2\tilde {p}_1^0 \cdot \tilde {p}_2^0 \cdot k^2\left( {\tilde
{p}_1^0 + p_1^0 + \tilde {p}_2^0 + p_2^0 } \right)}
\right]}{\tilde {p}_1^0 \cdot \tilde {p}_2^0 \left( {\tilde
{p}_1^0 + \tilde {p}_2^0 + p_1^0 - p_2^0 } \right){\kern 1pt}
{\kern 1pt} \left( {\tilde {p}_2^0 + \tilde {p}_1^0 - p_1^0 +
p_2^0 } \right){\kern 1pt} {\kern 1pt} \left( {\left( {\tilde
{p}_1^0 + p_1^0 } \right)^2 - k^2} \right){\kern 1pt} \left(
{\left( {\tilde {p}_2^0 + p_2^0 } \right)^2 - k^2}
\right)}\;{\kern 1pt}   \right\}.
 \end{array}
 \end{equation}

With the accuracy degree taken by us,
\begin{equation}
\label{eq4} \bar {u}\left( {p_2 } \right) \cdot \vec {\gamma
}{\kern 1pt} \vec {A} \cdot u\left( {p_1 } \right)\simeq u\ast
\left( {p_2 } \right)\left( {\frac{\vec {\sigma }{\kern 1pt} \vec
{H}}{2m} + \frac{\left( {\vec {p}_2 + \vec {p}_1 } \right)\vec
{A}}{2m}} \right)\,u\left( {p_1 } \right).
\end{equation}

The contribution to the anomalous magnetic moment is seen to be made only by
the first term.

In view of (\ref{eq4}), in the integrand of (4) we can restrict our consideration
only to those terms that do not depend on impulses $\vec {p}_1 ,\vec {p}_2
$. Then the integration in (4) reduces to one-dimensional integral in $k$. The
integration results can be written as
\begin{equation}
\label{eq5} \left( { - e} \right)A^i \cdot \Lambda ^{(3) i}(p_1
,p_2 ) \!= \!(e)\frac{e^2}{4\pi ^2}\frac{\vec {\sigma } \vec
{H}}{4m}\!\left[ {2\ln k \!- 3\ln \left| {k\! + \mu } \right|\! -
\frac{2}{3}\frac{k}{\mu } \!- \frac{1}{3}\frac{k\mu }{m^2} +
\frac{1}{3}\frac{k^2}{m^2}} \right]\mathop {\left|
{\begin{array}{l}
 \\
 \\
 \end{array}}
\right.}
 \limits_0^m.
\end{equation}

In (\ref{eq5}), $\mu = \left( {k^2 + m^2} \right)^{1 / 2}$.

\subsection{ Mass operator. Mass renormalization}

The contribution of the mass operator and electron mass
renormalization related counterterm to the electron anomalous
magnetic moment is described by diagrams $1b$, $1c$ for the
electron with impulse $p_{2}$ and diagrams $1d$, $1e$ for the
electron with impulse $p_{1}$. To estimate the contribution of
diagrams, for example, $1b$ and $1c$, use Heitler approach /2/ and
write the total process amplitude as
\begin{equation}
\label{eq6} G_3 \!= \!\frac{ie^2}{\left( {2\pi }
\right)^{\!4}}\bar {u}\left( {p_2 } \right)\left( { - e}
\right)\vec {\gamma } \vec {A} u\left( {p_1 } \right)\!\int
\!\!{d^{\;4}k\!\frac{2m^2 - 2p_2 \cdot k + k^2 -
\frac{2}{m^2}\left( {p_2 \cdot k} \right)^2}{\left( {k^0 \!- \!k}
\right) \!\left( {k^0 \!+\! k} \right)\! \left( {k^0 \!\!-\! p_2^0
- \!\tilde {p}_2^0 } \right)^2\!\left( {k^0 \!\!- \!p_2^0 \!+
\!\tilde {p}_2^0 } \right)^2}} .
\end{equation}
Then integrate (\ref{eq6}) with respect to ${\kern 1pt} k_0 $,
using the residue theorem with poles either in the right
half-plane of complex variable ${\kern 1pt} k_0 $, with the
profile line closure in the lower half-plane or with poles in the
left half-plane with the profile line closure in the upper
half-plane. Upon the integration with respect to ${\kern 1pt} k_0
$, with taking into account the accuracy degree taken by us, only
the terms independent on impulse $\vec {p}_2 $ can be left in the
integrand. On the integration with respect to the solid angle and
variable $k$ expression (\ref{eq6}) can be written as
\begin{equation}
\label{eq7}
G_3 = \bar {u}\left( {p_2 } \right)\,\left( { - e} \right)\frac{\vec {\sigma
}{\kern 1pt} \vec {H}}{2m}\,u\left( {p_1 } \right) \cdot \frac{e^2}{8\pi
^2}\left( {\ln k - \frac{3}{2}\ln \,\left| {k + \mu } \right| -
\frac{k^2}{2m^2} + \frac{k\mu }{2m^2}} \right)\;\mathop {\left|
{\begin{array}{l}
 \\
 \\
 \end{array}} \right.}\limits_0^m.
\end{equation}
Evidently, with the accuracy degree taken, the contribution of the
diagrams $1d$, $1e$ is also determined by relation (\ref{eq7}).

In view of (\ref{eq5}), the total contribution made by the
diagrams $1a\div 1e$ to the electron anomalous magnetic moment is
\begin{equation}
\label{eq8}
\begin{array}{l}
 \bar {u}\left( {p_2 } \right)\frac{\alpha }{4\pi  m}\,\left( { - e}
\right)\vec {\sigma }{\kern 1pt} \vec {H}\,\left( {\frac{2}{3}{\kern 1pt}
\frac{k}{\mu } - \frac{4}{3}\,\frac{k^2}{m^2} + \frac{4}{3}\,\frac{k\mu
}{m^2}} \right)\,\mathop {\left| {\begin{array}{l}
  \\
 \end{array}} \right.}\limits_0^m  u\left( {p_1 } \right) = \\
  = \bar {u}\left( {p_2 } \right)\frac{\alpha }{4\pi m}\left( { - e}
\right)\vec {\sigma } \vec {H}\left( {\frac{5}{3}\sqrt 2 -
\frac{4}{3}} \right)u\left( {p_1 } \right)\simeq\bar {u}\left(
{p_2 } \right)\frac{\alpha }{4\pi m}1.0237\left( { - e}
\right)\vec {\sigma }\vec {H} \cdot u\left( {p_1 } \right) .
 \end{array}
\end{equation}
In (\ref{eq8}), $\alpha = \frac{e^2}{4\pi }$ is the fine structure
constant.

\section{Lamb shift }

In this section we consider the electron motion in weak external
electric field $\vec {\varepsilon } = - \nabla A_0 $. In the final
expressions we will restrict our consideration to the terms
proportional to electric field $\varepsilon _i $ and expansions up
to the ones quadratic in electron impulse $p_i^2 $. Expressions
proportional to $p_i^2 \varepsilon _\kappa $ will be neglected.
The contribution to the Lamb shift is made by all diagrams $1a\div
1f$.

\subsection{Vertex operator}

Consider the contribution made by the diagram $1a$ with the vertex
operator $\Lambda _0^{(3)} (p_1 ,p_2 )$ to the Lamb shift. With
account for the accuracy degree taken by us, product $eA_0 \Lambda
_0^{(3)} (p_1 ,p{ }_2)$ can be written as
\begin{equation}
\label{eq9}
\begin{array}{l}
\hspace{-5mm}eA_0 \Lambda _0^{(3)} (p_1 ,p_2 ) = \\ = eA_0 ( -
)\frac{ie^2}{(2\pi )^{\;4}}\int {d^{\;4}\kappa \frac{B_1 + C_1
\cdot k_0 + D_1 \cdot k_0^2 }{(k_0 - k)(k_0 + k)(k_0 - p_1^0 -
\tilde {p}_1^0 )(k_0 - p_1^0 + \tilde {p}_1^0 )(k_0 - p_2^0 -
\tilde {p}_2^0 )(k_0 - p_2^0 + \tilde {p}_2^0 )}}.
 \end{array}
\end{equation}
In (\ref{eq9}), $B_1 = (4m^2 + 2\vec {q}^2 - 2k^2)\gamma _0 +
4i\gamma _0 \vec {\sigma }(\vec {k}\times \vec {q}),$ $C_1 = - 4m
+ 4\vec {\gamma }\,\vec {k}$, $D_1 = $\linebreak $=- 2\gamma _0.$

Next, like in Section 2, perform the integration with respect to
variable $k_0 $ and algebraic transformations with canceling from
the denominators the expression $(\tilde {p}_1^0 + p_1^0 - p_2^0 -
\tilde {p}_2^0 )$ that vanishes with $\vec {p}_1 ,\;\vec {p}_2 \to
0.$

As a result, expression (\ref{eq9}) can be written as:
\begin{equation}
\label{eq10}
\begin{array}{l}
 eA_0 \Lambda _0^{(3)} (p_1 ,p_2 )\! = \!- eA_0 \!\frac{e^2}{2(2\pi
 )^3}\int\!
{d\vec {k}\left\{ {\frac{(4m^2 + 2\vec {q}^2 - 2k^2)\gamma _0 -
4i\gamma _0 (\vec {\sigma }\times \vec {q})\vec {k} - 4m \cdot k +
4 \cdot \kappa \cdot \vec {\gamma }\vec {k} - 2\gamma _0 \cdot
k^2}{k(\tilde {p}_1^0 - k + p_1^0 )(\tilde {p}_1^0 + k - p_1^0
)(\tilde {p}_2^0 - k + p_2^0 )(\tilde {p}_2^0 + k - p_2^0 )}}
\right.} - \\[14pt]%
\quad - \frac{(4m^2 + 2\vec {q}^2 - 2k^2 - 4i(\vec {\sigma }\times
\vec {q}) \cdot \vec {k}) \cdot \gamma _0 [\tilde {p}_2^o ((\tilde
{p}_2^0 + p_2^0 )^2 - k^2) + \tilde {p}_1^0 ((\tilde {p}_1^0 +
p_1^0 )^2 - k^2) + 2\tilde {p}_1 \cdot \tilde {p}_2^0 (\tilde
{p}_1^0 + p_1^0 + \tilde {p}_2^0 + p_2^0 )]}{\tilde {p}_1^0 \tilde
{p}_2^0 (\tilde {p}_1^0 + \tilde {p}_2^0 + p_1^0 - p_2^0 )(\tilde
{p}_2^0 + \tilde {p}_1^0 - p_1^0 + p_2^0 )((\tilde {p}_1^0 + p_1^0
)^2 -
k^2)((\tilde {p}_2^0 + p_2^0 )^2 - k^2)} - \\[14pt]%
\quad - \frac{( - 4m + 4\vec {\gamma }\vec {k})[\tilde {p}_2^0
(\tilde {p}_1^0 + p_1^0 )((\tilde {p}_2^0 + p_2^0 )^2 - k^2) +
\tilde {p}_1^0 (\tilde {p}_2^0 + p_2^0 )((\tilde {p}_1^0 + p_1^0
)^2 - k^2) + 2\tilde {p}_1^0 \cdot \tilde {p}_2^0 ((\tilde {p}_1^0
+ p_1^0 )(\tilde {p}_2^0 + p_2^0 ) + k^2)]}{
 \tilde {p}_1^0 \cdot \tilde {p}_2^0 (\tilde {p}_1^0 + \tilde {p}_2^0 +
p_1^0 - p_2^0 )(\tilde {p}_2^0 + \tilde {p}_1^0 - p_1^0 + p_2^0
)((\tilde {p}_1^0 + p_1^0 )^2 - k^2)((\tilde {p}_2^0 + p_2^0 )^2 -
k^2)
 } + \\[14pt]%
\quad + \left. {\frac{2\gamma _0 \cdot [\tilde {p}_2^0 (\tilde
{p}_1^0 + p_1^0 )^2((\tilde {p}_2^0 + p_2^0 )^2 - k^2) + \tilde
{p}_1^0 (\tilde {p}_2^0 + p_2^0 )^2((\tilde {p}_1^0 + p_1^0 )^2 -
k^2) + 2\tilde {p}_1^0 \cdot \tilde {p}_2^0 \cdot k^2(\tilde
{p}_1^0 + \tilde {p}_2^0 + p_1^0 + p_2^0 )}{\tilde {p}_1^0 \cdot
\tilde {p}_2^0 (\tilde {p}_1^0 + \tilde {p}_2^0 + p_1^0 - p_2^0
)(\tilde {p}_2^0 + \tilde {p}_1^0 - p_1^0 + p_2^0 )((\tilde
{p}_1^0 + p_1^0 )^2 - k^2)((\tilde {p}_2^0 + p_2^0 )^2 - k^2)}}
\right\}.
 \end{array}
\end{equation}

If the integrand in (\ref{eq10}) is expanded in degrees of electron impulses $\vec
{p}_1 ,\;\vec {p}_2 $ up to the quadratic ones inclusive, then on the
averaging over angles integral (\ref{eq10}) reduces to the one-dimensional integral
in variable $k$.

The results of the integration of (\ref{eq10}) can be represented
as a sum of addends proportional to $\vec {\sigma }\left( {\nabla
A_{0\;} \times \vec {p}} \right),\quad \left( {\vec {p}_1 \cdot
\vec {p}_2 } \right)A_0 $ and $\left( {\vec {p}_1^{\;2} + \vec
{p}_2^{\;2} } \right)\;A_0 $.

The addends proportional to $\sigma \left( {\nabla A_0 \,\times
\vec {p}} \right)$ correspond to the contribution of the
interaction of the particle anomalous moment with external
electric field. The addends proportional to $\left( {\vec {p}_1
\cdot \vec {p}_2 } \right)A_0 $ and $ \left( {\vec {p}_1^{\;2} +
\vec {p}_2^{\;2} } \right)\;A_0 $ should be integrated into the
gauge-invariant expression proportional to $\nabla ^2A_0 =
$\linebreak $= - \left( {\vec {p}_1 - \vec {p}_2 } \right)^2  A_0
$.

When deriving the final expression for the integration of (\ref{eq10}) which is
proportional to $\vec {\sigma }\left( {\nabla A_{0\;} \times \vec {p}}
\right)$ with account for the accuracy degree taken by us, it is necessary
to include in the integrand expansion only the terms linear in impulses
$p_{1}$, $p_{2}$.

The final result of the integration of the terms proportional to $\vec
{\sigma }\left( {\nabla A_{0\;} \times \vec {p}} \right)$ can be represented
as
\begin{equation}
\label{eq11} \frac{\alpha }{\pi }\, \frac{e\vec \sigma \left(
\nabla A_0 \times \vec {p} \right)}{4m^2}\;\left[
{\frac{2}{3}\;\frac{k\mu }{m^2}\, -
\,\frac{2}{3}\;\frac{k^2}{m^2}\, + \,\frac{2}{3}\;\frac{k^3}{\mu
^3}} \right]\;\mathop {\left| {\begin{array}{l}
 \\
 \\
 \end{array}} \right.}\limits_0^m.
\end{equation}

The integration of the terms proportional to $\left( {\vec {p}_1
\vec {p}_2 } \right)A_0 $ and $\left( {\vec {p}_1^{\;2} + \vec
{p}_2^{\;2} } \right) A_0 $ entails, like the conventional QED
calculations do, the problem of the logarithmic divergence in the
lower limit of integration with $k \to 0$. To overcome this
problem, divide the region of integration into two parts:
low-energy region $(0 \le k < \gamma \, m)$ and high-energy region
$(\gamma  m \le k \le \,m)$, where $\gamma \ll 1$.

When integrating over the low-energy region, we can make use of
the results of the nonrelativistic perturbation theory (see, e.g.,
/3/). Then for the terms proportional to $\left( {\vec {p}_1 \vec
{p}_2 } \right)A_0 $ the result of the integration in (\ref{eq10})
can be written as
\begin{equation}
\label{eq12} \left[ {eA_0 \cdot \Lambda _0^{(3)} \,\left( {p_1
,p_2 } \right)} \right]{\kern 1pt} _1 \, = \frac{2\left( {\vec
{p}_1 \cdot \vec {p}_2 } \right)\,A_0 }{m^2}\left( { -
\frac{\alpha }{\pi }} \right)\,\frac{1}{3}\,\ln \left(
{\frac{M}{\gamma \, m}} \right).
\end{equation}

In (\ref{eq12}), $M$ is some average electron bound state energy
level difference. The $M$ is evaluated numerically. For example,
$\ln \frac{m}{M} = 7.6876$ for the hydrogen atom energy level
shift with $n = 2$ /2/.

In the integration of (\ref{eq10}) over the high-energy region the
following is obtained for the terms proportional to $\left( {\vec
{p}_1 \cdot \vec {p}_2 } \right)  A_0 $:
\begin{equation}
\label{eq13}
\begin{array}{ll}
 \left[ {e\,A_0  \Lambda _0^{(3)} \left( {p_1 ,p_2 } \right)}
\right]_2 &\hspace{-3mm}= \frac{2 \left( {\vec {p}_1 \vec {p}_2 }
\right) e A_0 }{m^2}\left( { - \frac{\alpha }{\pi }} \right)\left[
{\frac{1}{3}\ln \left( {1 + \frac{\mu }{k}} \right) -
\frac{1}{4}\frac{k}{\mu } - \frac{1}{36}\frac{k^3}{\mu ^3}}
\right]\left. {\begin{array}{l}
 \\
 \\
 \end{array}} \right|_{\gamma \, m}^m = \\[5mm]
& \hspace{-3mm}= \frac{2 \left( {\vec {p}_1 \vec {p}_2 } \right)
\left( {e A_0 } \right)}{m^2}\left( { - \frac{\alpha }{\pi }}
\right)\left[ { - \frac{1}{3}\,\ln \frac{1}{\gamma } +
\frac{1}{3}\ln \left( {1 + \sqrt[]{2}} \right) - \frac{\sqrt 2
}{8} - \frac{\sqrt 2 }{36 \cdot 4}} \right] .
 \end{array}
\end{equation}

The dependence on $\gamma $ is seen to disappear in the summation of (\ref{eq12})
and (\ref{eq13}), which validates the choice of the low-energy and high-energy
regions of integration of (\ref{eq10}).

The calculation of the terms proportional to $(p_1^2 + p_2^2 )A_0
$ requires, besides the integration of the relevant parts of
(\ref{eq10}), inclusion of the contribution made by the diagrams
$1â\div 1e$. The results of the integration are presented in the
following section.

\subsection{Mass operator. Mass renormalization}

The contribution of the mass operator and electron mass
renormalization related counterterm to the Lamb shift is described
by the diagrams 1b, 1c for the electron with impulse $p_{2}$ and
the diagrams $1d$, $1e$ for the electron with impulse $p_{1}$. The
contribution made by the diagrams $1b$, $1c$ is determined by
relation (\ref{eq6}) with substitution $ - e\,\vec {\gamma }\vec
{A} \to \,e\gamma _0 A_0 $ in the bispinor coverings. On the
substitution and integration with respect to $k_0 $ expression
(\ref{eq6}) can be written as
\begin{equation}
\label{eq14}
\begin{array}{l}
 G_3 = \frac{e^2}{(2\pi )^3}\;\bar u(p_2 )e\gamma _0 A_{0} u(p_1 )\int
{d \vec k }\left\{ \frac{2m^2 - 2p_2^0 k + 2 p_2 \vec k- \frac
2{m^2}{\left( p_2^0  k - \vec p_2 \vec k  \right)}^2}{2k{\left( k
- p_2^0 - \tilde {p}_2^0  \right)}^2{\left( k - p_2^0 + \tilde
{p}_2^0  \right)}^2} \;+  \right. \\[15pt]%
 \qquad \;+ \;\frac{2\tilde
{p}_2^0 - \frac{4}{m^2}\,\left( {p_2^0 } \right)^2\,\left( {p_2^0
+ \tilde {p}_2^0 } \right) + \frac{4}{m^{2}}p_2^0 \left( {\vec
{p}_2 \vec {k}} \right)}{\left( {\tilde {p}_2^0 + p_2^0 - k}
\right)\;\left( {\tilde {p}_2^0 + p_2^0 + k} \right) \;\cdot
4\left(
{\tilde {p}_2^0 } \right)^2} \;- \\[17pt]%
 \qquad \;- \!\left( 2m^2 \!- 2\!\left(
{p_2^0 } \right)^2 \!- 2p_2^0 \tilde {p}_2^0 \!+ 2\vec p_2 \vec k
+ \left( \tilde p_2^0 \!+ p_2^0  \right)^2 \!- k^2 -\!
\frac{2}{m^2}\left( {p_2^0 } \right)^2\left(
p_2^0 \!+ \tilde {p}_2^0 \right)^2 \!+\right. \\[9pt]%
 \qquad \;+
\left.\frac{4}{m^2}p_2^0 \left( p_2^0 + \tilde {p}_2^0
\right)\left( \vec p_2 \vec k\right) - \frac{2}{m^2}{\left( \vec
{p}_2 \vec {k} \right)}^2\right) \left[ \frac{1}{4\left( {\tilde
{p}_2^0 } \right)^2\,\left( {\tilde {p}_2^0 + p_2^0 - k}
\right)^2\,\left( {\tilde {p}_2^0 + p_2^0 + k} \right)}\right. +
\\[15pt]%
\qquad\; +\left.\left. \frac{1}{4\left( {\tilde {p}_2^0 }
\right)^2\,\left( {\tilde {p}_2^0 + p_2^0 + k} \right)^2\,\left(
{\tilde {p}_2^0 + p_2^0 - k} \right)} + \frac{1}{4\left( {\tilde
{p}_2^0 } \right)^3\,\left( {\tilde {p}_2^0 + p_2^0 - k}
\right)\,\left( {\tilde {p}_2^0 + p_2^0 + k} \right)}\right]
\right\} .
 \end{array}
\end{equation}

It is evident that the contribution made by the diagrams 1$d, $1$e$ is also
determined by expression (\ref{eq14}) with substitution $p_2 \to p_1 $. To integrate
(\ref{eq14}) further, it is necessary to expand the integrand in the electron
impulse degrees up to the quadratic inclusive and, on the averaging over
angles, reduce to the one-dimensional integral with respect to variable $k$.

Above all we note that on the integration of the terms in (\ref{eq10}), (\ref{eq14}), and in
relation (\ref{eq14}) with substitution $p_2 \to p_1 $ proportional to $eA_0 $ their
total contribution is zero.

To integrate the terms in (\ref{eq10}) that are proportional to $
\left( {\vec {p}_1^{\;2} + \vec {p}_2^{\;2} } \right)A_0 $, like
previously, divide the region of integration into two parts in
expression (\ref{eq14}) and in expression (\ref{eq14}) with
substitution $p_2 \to p_1 $: low-energy region $\left( {0 \le k <
\gamma \; m} \right)$ and high-energy region $\gamma \; m \le k
\le m$, where $\gamma \ll 1$.

The result of the integration over the low-energy region of integration can
be shown to be
\begin{equation}
\label{eq15} \left[ {\bar {u}(p_2 )  \Lambda _0^{(3)} (p_1 ,p_2 )
eA_0 u(p_1 ) + G_3 } \right] _1 = \bar {u}(p_2 )\left( {\frac{\vec
{p}_1^{\;2} + \vec {p}_2^{\;2} }{m^2}} \right)eA_0 \frac{\alpha
}{\pi }\frac{1}{3}\ln \left( {\frac{M}{\gamma \; m}} \right) u(p_1
).
\end{equation}
The result of the integration over the high-energy region of
integration can be represented as
\begin{equation}
\label{eq16}
\begin{array}{l}
\hspace{-3mm} \left[ {\bar {u}(p_2 )  \Lambda _0^{(3)} (p_1 ,p_2 )
eA_0
 \;u(p_1 ) + G_3 } \right] _2 = \bar {u}(p_2 )\left\{
\left( {\frac{\vec {p}_1^{\;2} + \vec {p}_2^{\;2} }{m^2}}
\right)\,eA_0 \cdot \frac{\alpha }{\pi }\left[ \frac{1}{3}\ln
\left( {1 \!+ \frac{\mu }{k}} \right) - \right.\right.\\[9pt]%
\qquad\left.\left. - \frac{7}{12}\,\frac{k}{\mu } \, -
\;\,\frac{1}{36}\;\frac{k^3}{\mu ^3}\,\; -\left.
\,\,\frac{1}{6}\;\frac{k^2}{m^2} + \frac{1}{6}\;\frac{k}{m} +
\frac{1}{6}\;\frac{k \cdot \mu }{m^2} \right]\; \right|_{\gamma
\cdot m}^m \, \right\} \;u(p_1 ) = \\%
 \qquad \,= \bar {u}(p_2
)\;\left\{ \frac{\left( \vec {p}_1^{\;2} + \vec {p}_2^{\;2}
\right) \; eA_0 }{m^2} \;\frac{\alpha }{\pi }\left[ { -
\frac{1}{3}} \ln \,\frac{1}{\gamma } + \frac{1}{3}\,\ln \,\left(
{1 + \sqrt 2 } \right) - \frac{\sqrt 2 }{8} - \frac{\sqrt 2 }{36
\cdot \,4} \right] \right\} u(p_1 ).
 \end{array}
\end{equation}

Like for expressions (\ref{eq12}), (\ref{eq13}), the dependence on
$\gamma $ disappears in summation (\ref{eq15}), (\ref{eq16}).
Moreover, the simultaneous summation of expressions (\ref{eq12}),
(\ref{eq13}), (\ref{eq15}), (\ref{eq16}) with the upper limit of
integration $|\vec {L}|= m$ allows their integration into a single
gauge-invariant expression proportional to $(\vec {p}_1 - \vec
{p}_2 )^{\;2}A_0 = - \nabla ^2A_0 $.
\subsection{Polarization operator. Charge renormalization}

Now consider the contribution made by the diagram $1f$. The
process amplitude can be written as
\[
\bar {u}(p_2 )\frac{(i)}{q^2}\left( {\gamma _0 \Pi_{00}^{(2)} (q)
-
 \gamma _i \Pi_{i0}^{(2)} (q)} \right)\,eA_0 (q)u(p_1 ),
\]
\noindent where $\Pi_{\mu \nu }^{(2)} (q)$ is the polarization
operator.
\begin{equation}
\label{eq17} \Pi_{\mu \nu }^{(2)} (q) = \frac{4e^2}{(2\pi
)^{\;4}}\int {d^{\;4}p\frac{\delta _{\mu \nu }  ( - p^2 + m^2 +
pq) + 2p_\mu p_\nu - (p_\mu  q_\nu + p_\nu  q_\mu )}{(p^2 -
m^2)[(p - q)^2 - m^2]}}.
\end{equation}
The polarization operator component
\begin{equation}
\label{eq18} \Pi_{00}^{(2)} (q) = \frac{4e^2}{(2\pi )^{\;4}}\int
{d^{\;4}p\frac{(p_0^2 + E^2 - p_0 q_0 - \vec {p}\vec {q}\,)}{(p_0
- E)(p_0 + E)(p_0 - q_0 - \tilde {E})(p_0 - q_0 + \tilde {E})}}
\end{equation}
\noindent alone makes the contribution to the process amplitude in
the electron scattering in static external field $A_0 (\vec {x})$.
In (\ref{eq18}), like previously,$E = (m^2 + p^2)^{1 / 2}$;
$\tilde {E} = (m^2 + (\vec {p} - \vec {q})^2)^{1 / 2}$.

On integration of (\ref{eq18}) with respect to $dp_0 $ using the
residue theorem and keeping in mind that $q_0 = 0$ in the case
under discussion, we obtain
\begin{equation}
\label{eq19}
\begin{array}{ll}
 \Pi_{00}^{(2)} (q) &= \frac{4e^2}{(2\pi )^{\;4}}\frac{( - 2\pi i)}{2}\int {d\vec
{p}\left[ {\frac{2E^2 - \vec {p}\vec {q}}{E(E^2 - \tilde {E}^2)} +
\frac{\tilde {E}^2 + E^2 - \vec {p}\vec {q}}{\tilde {E}(\tilde
{E}^2 - E^2)}} \right]} = \\[7pt]%
& = \frac{4e^2}{(2\pi )^{\;4}}( - \pi i)\int {d\vec {p}\left[
{\frac{E(\tilde {E} - E) + \vec {p}\vec {q}}{\tilde {E} \cdot
E(\tilde {E} + E)}} \right]}.
 \end{array}
\end{equation}

Expanding the integrand in (\ref{eq19}) in powers of $q^{i}$ up to
$\vec {q}^{\;4}$ and integrating, we arrive at
\begin{equation}
\label{eq20}
\begin{array}{ll}
 \Pi_{00}^{(2)} (q) =& \vec {q}^{\;2}\left( { - \frac{2i\alpha }{\pi }}
\right)\;\left[ {\frac{1}{3}\ln (p + p_0 ) -
\frac{1}{3}\frac{p}{p_0 } + \frac{1}{18}\frac{p^3}{p_0^3 }}
\right]\mathop {\left| {\begin{array}{l}
 \\
  \end{array}} \right.}\limits_0^{\left| \vec {L} \right|} +
  \\[12pt]
& + \vec {q}^{\;4}\left( { - \frac{i\alpha }{\pi }}
\right)\;\left[ { - \frac{1}{3}\frac{p^3}{\tilde {p}_0^3 } +
\frac{5}{12}\frac{p^5}{\tilde {p}_0^5 } -
\frac{3}{20}\frac{p^7}{\tilde {p}_0^7 }} \right]\mathop {\left|
{\begin{array}{l}
 \\
  \end{array}} \right.}\limits_0^{\left| \vec {L} \right|} .
 \end{array}
\end{equation}

The first addend in (\ref{eq20}) requires the charge normalization, whereas the
second addend contributes to the Lamb energy level shift.

At the moment the author has no strong arguments in favor of the finiteness
of the upper limit of integration for the polarization operator other than
the speculations that the limiting impulses $L$ for the mass and polarization
operators are of the same magnitude.

Hence, in this paper we use $\Pi_{00}^{(2)} (q)$ with $\left| \vec
{L} \right| \to \infty $. In this case we obtain the same results
as in the conventional quantum electrodynamics calculations.
\begin{equation}
\label{eq21} \Pi_{00}^{(2)} (q) = \vec {q}^{\;2}\left( { -
\frac{i\alpha }{\pi }} \right)\left[ {\frac{2}{3}\ln \frac{\left|
\vec {L} \right|}{m} - \frac{5}{9}} \right] + \vec {q}^{\;4}\left(
{\frac{i\alpha }{\pi }} \right) \cdot \frac{1}{15}
\end{equation}

\section{Discussion of results}

Initially write Dirac Hamiltonian for the electron motion in weak external
electromagnetic field.
\begin{equation}
\label{eq22} H_D \approx m + \frac{\left( {\vec {p} - e\vec {A}}
\right)^2}{2m} + eA_0 - \left( {\frac{e\vec {\sigma }}{2m}}
\right)\vec {H} - \frac{e\vec {\sigma }}{2m}\left( {\frac{\vec
{\varepsilon }\times \vec {p}}{2m}} \right) - \frac{e}{8m^2}\nabla
\vec {\varepsilon }.
\end{equation}
Pay attention to the third and fourth terms, which represent the
interaction of electron magnetic moment $\vec {\mu } = -
\frac{e}{2m}\vec {\sigma }$ with the magnetic and electric fields.

The existence of the intrinsic electromagnetic field in the electron leads
to radiation corrections, a part of which has been calculated in this paper
in the second order of the perturbation theory by the algorithm discussed in
/1/ and in Introduction. These results are presented below in comparison
with the conventional quantum electrodynamics calculations with infinite
limits of integration.

\subsection{Anomalous magnetic moment}

One of the radiation corrections in the interaction of electron with
external static magnetic field is appearance of an additional term in the
interaction energy which is identified with the electron magnetic moment
additional in comparison with the Dirac one.

In the conventional QED calculations in the second order of the perturbation
theory it is
\begin{equation}
\label{eq23} \vec {\mu } = \vec {\mu }_D + \Delta \vec {\mu
}^{(2)} = - \frac{e\vec {\sigma }}{2m}\left( {1 + \frac{\alpha
}{2\pi }} \right),
\end{equation}
\noindent where $\Delta \vec {\mu }^{\;(2)} = - \frac{e\vec
{\sigma }}{2m}\frac{\alpha }{2\pi }$ is the anomalous magnetic
moment of electron.

In this paper, according to (\ref{eq8}),
\begin{equation}
\label{eq24}
\begin{array}{ll}
 \Delta \vec {\mu }^{(2)} &= - \frac{e\vec {\sigma }}{2m}\frac{\alpha }{2\pi
}\left( {\frac{2}{3}\,\frac{k}{\left( {k^2 + m^2} \right)^{1 / 2}}
- \frac{4}{3}{\kern 1pt} \frac{k^2}{m^2} + \frac{4}{3}{\kern 1pt}
\frac{(k^2 + m^2)^{1 / 2}}{m^2}} \right)\mathop {\left|
{\begin{array}{l}
 \\
 \\
 \end{array}} \right.}\limits_0^m = \\
& = - \frac{e\vec {\sigma }}{2m}\frac{\alpha }{2\pi }\left(
{\frac{5}{3}\sqrt 2 - \frac{4}{3}} \right) \approx - \frac{e\vec
{\sigma }}{2m}\frac{\alpha }{2\pi }\cdot 1.0237.
 \end{array}
\end{equation}

\subsection{Lamb atomic energy level shift }

The radiation corrections in the second order of the perturbation theory to
Hamiltonian (\ref{eq22}) in the electron motion in weak external Coulomb field are a
sum of terms proportional to $\frac{\vec {\varepsilon }\times \vec
{p}}{2m^2},\;\frac{\left( {p_1^2 + p_2^2 } \right)A_0 }{m^2},\,\frac{2\left(
{\vec {p}_1 \vec {p}_2 } \right)A_0 }{m^2}.$ In the conventional QED
calculations, the last two terms appear in the final results as
gauge-invariant expression $ - \frac{\left( {\vec {p}_1 - \vec {p}_2 }
\right)^2A_0 }{m^2} = \frac{\nabla ^2A_0 }{m^2}$.

First consider the term proportional to $\frac{\vec {\varepsilon }\times
\vec {p}}{2m^2}$. This is the radiation correction due to the electron's
anomalous magnetic moment, which interacts with the external electric field
in this case. The correction according to (\ref{eq22}) would seem to be
\[
\Delta \vec {\mu }^{(2)}\left( {\frac{\vec {\varepsilon }\times
\vec {p}}{2m}} \right) = - \frac{e\vec {\sigma
}}{2m}\,\frac{\alpha }{2\pi } \, \frac{\vec {\varepsilon }\times
\vec {p}}{2m}.
\]
However, in the conventional QED calculations the correction
proves two times as large,
\begin{equation}
\label{eq25} \left( {\Delta H_D } \right)_1 = - \frac{e\vec
{\sigma }}{2m}\,\,\frac{\alpha }{\pi }\,\frac{\vec {\varepsilon
}\times \vec {p}}{2m}.
\end{equation}
In the calculations of this paper, according to (\ref{eq11}), the
correction is
\begin{equation}
\label{eq26}
\begin{array}{l}
 \left( {\Delta H_D } \right)_1 = - \frac{e\vec {\sigma }}{2m}\,\frac{\alpha
}{2\pi }\left[ {\frac{4}{3}\frac{k(m^2 + k^2)^{1 / 2}}{m^2} -
\frac{4}{3}\frac{k^2}{m^2} + \frac{4}{3}\frac{k^3}{(m^2 + k^2)^{3 / 2}}}
\right]\;\mathop {\left| {\begin{array}{l}
 \\
 \\
 \end{array}} \right.}\limits_0^m  \left( {\frac{\vec {\varepsilon
}\times \vec {p}}{2m}} \right) = \\
 = - \frac{e\vec {\sigma }}{2m}\,\frac{\alpha }{2\pi }\left(
{\frac{5}{3}\sqrt 2 - \frac{4}{3}} \right)\left( {\frac{\vec
{\varepsilon }\times \vec {p}}{2m}} \right) \approx - \frac{e\vec
{\sigma }}{2m}\,\frac{\alpha }{2\pi } \cdot 1.0237 \; \frac{\vec
{\varepsilon }\times \vec {p}}{2m} = \Delta \vec {\mu }^{\;(2)}
\frac{\vec {\varepsilon }\times \vec {p}}{2m}.
 \end{array}
\end{equation}

From (\ref{eq26}) it is seen that the calculations according to
the algorithm of this paper yield symmetry of the interaction of
the anomalous magnetic moment with the magnetic and electric
field. True, correction (\ref{eq26}) is therewith $\sim 2$ times
as small as correction (\ref{eq25}) found from the conventional
QED calculations.

From the standpoint of the internal logic of QED, the results of
the calculations by the algorithm of this paper are more
preferable. Because of the presence of the intrinsic field in the
electron the magnetic moment changes:

\noindent$\vec {\mu } = \frac{e\vec {\sigma }}{2m}\left( {1 +
\frac{\alpha }{2\pi } \cdot 1.0237} \right);$ it is with this
changed magnetic moment that the electron interacts with electric
field $\left( {\frac{e\vec {\sigma }}{2m}\left( {1 + \frac{\alpha
}{2\pi } \cdot 1.0237} \right)\frac{\vec {\varepsilon }\times \vec
{p}}{2m}} \right)$, which leads to the atomic energy level shift.

The comparison with the associated experimental data will be dealt with
later.

We now turn to the terms proportional to $\left( {\vec {p}_1^{\;2}
+ \vec {p}_2^{\;2} } \right) A_0 $ and $(\vec {p}_1^{\;2} \cdot
\vec {p}_2^{\;2} )  A_0 $.

As already mentioned above, in the conventional QED calculations
with infinite limits of integration these terms are grouped into
the gauge-invariant expression proportional to $\nabla ^2A_0 $.
\begin{equation}
\label{eq27} \left( {\Delta H_D } \right)_2 = -
\frac{1}{8m^2}\frac{\alpha }{\pi } \cdot \frac{8}{3}\,\left( {\ln
\frac{m}{2M} + \frac{5}{6}} \right) \,\nabla \vec {\varepsilon }.
\end{equation}
In the calculations of this paper, this correction, according to
(\ref{eq12}), (\ref{eq13}), (\ref{eq15}), (\ref{eq16}), is
\begin{equation}
\label{eq28}
\begin{array}{l}
 \left( {\Delta H_D } \right)_2\! = \!\frac{2\vec {p}_1 \vec {p}_2 }{m^2}eA_0
\!\left( {\! - \frac{\alpha }{\pi }} \!\right)\!\!\left\{\!
{\frac{1}{3}\ln \frac{M}{m} \!+\! \frac{1}{3}\ln \!\left(\! {1
\!\!+\! \sqrt 2 } \right) \!+\! \!\left[\! { -
\frac{1}{4}\frac{k}{(k^2 \!+\! m^2)^{1 / \!2}} \!-\!
\frac{1}{36}\frac{k^3}{(k^2 \!+\! m^2)^{3\! /\! 2}}}
\right]\!\mathop {\left| {\begin{array}{l}
  \\
 \end{array}} \right.}\limits_0^m } \!\!\!\right\} \!\!+ \\[12pt]
  \quad +\frac{\left( {\vec {p}_1^{\;2}\! +\! \vec {p}_2^{\;2} } \right)}{m^2}eA_0 \!
 \left(\!
{\frac{\alpha }{\pi }} \right)\left\{\frac{1}{3}\ln \frac{M}{m} +
\frac{1}{3}\ln \!\left( {1 + \sqrt 2 } \right) + \left[
 - \frac{7}{12}\frac{k}{(k^2 + m^2)^{1 / 2}} -
\frac{1}{36}\frac{k^3}{(k^2 + m^2)^{3 / 2}} -
\right.\right.\\[7pt]
  \quad -\left.\left.\frac{1}{6}\frac{k^2}{m^2} + \frac{1}{6}\frac{k}{m} + \frac{1}{6}k(k^2 +
m^2)^{1 / 2} \right]\!\mathop {\left| {\begin{array}{l}
  \\
 \end{array}} \right.}\limits_0^m \! \right\} .
 \end{array}
\end{equation}

From (\ref{eq28}) it is seen that it is only with the upper limit
of integration $|\vec {L}|=m$ that the terms can be grouped into
the gauge-invariant expression proportional to $(\vec {p}_1 - \vec
{p}_2 )^{\;2}eA_0 = - \nabla ^2(eA_0 )$. In this case
\begin{equation}
\label{eq29} \left( {\Delta H_D } \right)_2 = \frac{ - \nabla
^2(eA_0 )}{m^2}\frac{\alpha }{\pi }\left\{ {\frac{1}{3}\ln
\frac{M}{m} + \frac{1}{3}\ln (1 + \sqrt 2 ) - \frac{19}{144}\sqrt
2 } \right\}.
\end{equation}

With allowance made for the correction due to the contribution of
the diagram $1f $ (vacuum polarization), the final expression for
the radiation correction in the second order of the perturbation
theory leading to the Lamb energy level shift can be written as
follows:
\begin{itemize}
  \item for the conventional QED calculations:
\begin{equation}
\label{eq30} \Delta H_D = - \frac{e\vec {\sigma }}{2m}\frac{\alpha
}{\pi }\frac{\vec {\varepsilon }\times \vec {p}}{2m}\, -
\,\frac{1}{3m^2}\;\frac{\alpha }{\pi }\left( {\ln \frac{m}{2M} +
\frac{5}{6} - \frac{1}{5}} \right)\nabla \vec {\varepsilon };
\end{equation}
  \item for the calculations of this paper with the finite limit
of integration $ | \vec {L}|=m$:
\begin{equation}
\label{eq31} \hspace{-5mm}\Delta H_D\! = - \frac{e\vec {\sigma
}}{2m}\frac{\alpha }{2\pi }\cdot 1.0237  \frac{\vec {\varepsilon
}\!\times \!\vec {p}}{2m} - \frac{1}{3m^2}\frac{\alpha }{\pi
}\!\left(\! {\ln\! \frac{m}{M} - \ln \!\left( \!{1 \!+ \!\sqrt 2 }
\right)\! + \frac{57}{144}\sqrt 2 \!-\! \frac{1}{5}}
\right)\!\nabla \vec {\varepsilon }.
\end{equation}
\end{itemize}

The term in (\ref{eq31}) which is proportional to $\nabla \vec {\varepsilon }$ makes
about 0.94 of the associated term in (\ref{eq30}) derived from the conventional QED
calculations. The Lamb hydrogen level shift $\Delta \nu _{2S_{1 / 2} } -
\Delta \nu _{2P_{1 / 2} } $ calculated from relation (\ref{eq31}) is also smaller by
$\sim $6{\%} than that calculated from (\ref{eq30}).

Summarize the results of ref. /1/ and this paper.

The merits of the suggested relativistically- and gauge-invariant algorithm
for the self-energy expression regularization in quantum electrodynamics
are:
\begin{itemize}
  \item possibility of finite renormalization of electron and positron mass /1/;
  \item agreement between the final results for the self-energy in the second order
of the ``old'' and relativistically covariant perturbation theory
/1/;
  \item provision of the symmetric interaction of the electron intrinsic magnetic
moment with the external magnetic and electric fields.
\end{itemize}

What gives us concern is the disagreement between the calculations
in the second order of the perturbation theory and the
conventional QED calculations in regard to the electron anomalous
magnetic moment (larger by $\sim 2.4\,{\%}$ in the former case)
and Lamb hydrogen level shift $\Delta \nu _{2S_{1 / 2} } - \Delta
\nu _{2P_{1 / 2} } $ (smaller by $\sim 6\,{\%}$).

The answer to the question of the agreement between the experimental data
and results of the calculations by the algorithm of this paper will be given
by the calculations of the next order of the perturbation theory that are
being planned for the nearest future.

For the anomalous magnetic moment calculation in the next order of the
perturbation theory to agree with the experimental data, the following
result should be obtained.
\begin{equation}
\label{eq32} \vec {\mu } = - \frac{e\vec {\sigma }}{2m}\left( {1 +
\frac{\alpha }{2\pi } \cdot 1.0237 - 4.77\left( {\frac{\alpha
}{\pi }} \right)^2} \right).
\end{equation}

In the QED calculations with infinite limits of integration the
coefficient $\alpha _4 $ of $\left( {\frac{\alpha }{\pi }}
\right)^2$ is /4/, /5/, /6/
\begin{equation}
\label{eq33} \alpha _4 =  0.3285  =  1.3681 + 0.8222 -  6.8371 +
4.9753.
\end{equation}

It is seen to be the algebraic sum of quite large numbers, and it
is not at all impossible that the integration by the algorithm of
this paper can yield $\alpha _4  \simeq  -4.77$.

Similarly, for the calculation of the Lamb shift of hydrogen
levels $2S_{1 / 2}$ and $2P_{1 / 2}$ by the algorithm of this
paper with $|\vec {L}|= m$ with account for three orders of the
perturbation theory to agree with the experimental data, for the
last term in (\ref{eq22}) proportional to $\nabla \vec
{\varepsilon }$ the result should be as follows:
\begin{equation}
\label{eq34}
 - \frac{\nabla \vec {\varepsilon }}{8m^2}\left( {1 + 20.874\frac{\alpha
}{\pi }\,\left( {0.94 + 28.6\frac{\alpha }{\pi }} \right)}
\right).
\end{equation}

Thus, the final judgement about the applicability of the algorithm
discussed in this paper for the calculation of the radiation
corrections in quantum electrodynamics should be provided by
higher-order perturbation theory calculations.

\addcontentsline{toc}{section*}{\hspace{-6mm}References \hfill}
%\begin{thebibliography}{99}
\section*{References}
\begin{enumerate}
\item {\em Gichuk A.V., Neznamov V.P., Petrov Yu.V.} Feasibility
of finite renormalization of particle mass in quantum
electrodynamics, hep-th/0301245.
\item {\em Heitler V.} Quantum
theory of radiations. Moscow, Inostrannaya Literature Publishers,
1956.
\item {\em Akhiezer A.N., Berestetsky V.V.} Quantum
electrodynamics. Moscow, Nauka Publishers. Editorial Office of
Physical and Mathematical Literature. 1969.
\item {\em Karplus
R., Krull N.M.} Phys.Rev. 1950. V.77. P.536.
\item {\em
Sommerfeld A.} Phys.Rev. 1957. V.107. P.328.
\item {\em
Petermann} Helv.Phys.Acta. 1957. V.30. P.407.
\end{enumerate}
\end{document}